\documentclass[aps,prb,twocolumn,floatfix,amssymb]{revtex4}

\newcommand{\bn}[1]{\mbox{\boldmath $#1$}}

\usepackage{amsmath}
\usepackage{amsxtra}
\usepackage{graphicx}

\begin{document}

\title{Spin Hall and longitudinal conductivity of a conserved spin current in two dimensional heavy-hole gases} 
\author{Arturo Wong{\email{wong@phy.ohiou.edu}} and Francisco Mireles}  
           
\affiliation{Dpto. de F{\'\i}sica Te\'orica, Centro 
de Nanociencias y Nanotecnolog{\'\i}a, Universidad Nacional 
Aut\'onoma de M\'exico, Apdo. Postal 2681, 22800 Ensenada, 
Baja California, M\'exico.}
               
\date{\today}          
  
\begin{abstract}

The spin Hall and longitudinal conductivity of a 2D heavy-hole gas with {\it k}-cubic Rashba and Dresselhaus spin-orbit interaction is studied in the ac frequency domain. Using Kubo linear-response theory and a recently proposed definition for the (conserved) spin current operator suitable for spin-3/2 holes, it is shown that the spin conductivity tensor exhibit very distinguishable features from those obtained with the standard definition of the spin current. This is due to a significant contribution of the spin-torque term arisen from the alternative definition of spin current which strongly affects the magnitude and the sign of the dynamic spin current. In the dc (free of disorder) limit, the spin Hall conductivity for only (or dominant) {\it k}-cubic Rashba coupling is $\sigma^{s,z}_{xy}(0)=-9e/8\pi$, whereas $\sigma^{s,z}_{xy}(0)=-3e/8\pi$ for only (or dominant)  {\it k}-cubic Dresselhaus  coupling. Such anisotropic response is understood in terms of the absence of mapping the {\it k}-cubic Rashba $\leftrightarrow$ Dresselhaus Hamiltonians. This asymmetry is also responsible for the non-vanishing dc spin Hall conductivity ($\sigma^{s,z}_{xy}(0)=-6e/8\pi$)  when the  Rashba and Dresselhaus parameters have the same strength, in contrast with its corresponding case for electrons. These results are of  relevance to validate the alternative definition of spin-current through measurements in the frequency domain of the spin accumulation and/or spin currents in 2D hole gases.

\end{abstract}

\pacs{}

\maketitle

\section{INTRODUCTION}

Nowadays spintronics research is devoted to great extent to the study of fundamental issues concerning how to generate, manipulate and detect spin currents in a controlled and efficient way.\cite{Zutic} In particular, spin currents can be generated in solid state systems through the spin Hall effect (SHE),\cite{Dyakonov,Hirsch,Zhang}  a relativistic spin-orbit interaction (SOI) phenomenon that converts an electrical voltage into a pure spin current. The SHE is a result of a spin-dependent  scattering from charged impurities through side jump and/or skew processes \cite{Hankiewicz-Vignale} due to spin-orbit coupling (extrinsic SHE)\cite{Dyakonov,Hirsch}. It can be generated also in bulk and heterostructure due to built-in fields that modifies the band structure via the SOI (intrinsic SHE).\cite{Murakami, Sinova}

The SHE has been inferred by optical means via the spin Hall accumulation at the edges of a doped ({\it n}-type) semiconductor channel\cite{Kato,Stern}, as well as in two-dimensional ({\it p}-type) hole gases\cite{Wunderlich} (2DHGs). It has been also detected electrically in metals  throughout its reciprocal effect, the inverse SHE,\cite{Saitoh,Kimura,Valenzuela}  -- the generation of a charge current by a transverse spin-current in materials with SOI. \cite{Ando} Recently evidence of a rather large spin Hall signal (2.9 m$\Omega$) at room temperature in FePt/Au multiterminal devices exhibiting significant skew scattering processes has been also observed.\cite{Seki} Most recently, Br\"une {\it et al}.\cite{Bruene} have reported the first electrical measurement and manipulation of the intrinsic SHE using ballistic HgTe semiconductor nanostructures observing even larger spin Hall signals of the order of few $k\Omega$.

Theoretically, the intrinsic dc spin Hall conductivity in 2D electron systems with $k$-linear Rashba and/or Dresselhaus type of SOI is known to be suppressed by weak (non magnetic) disorder  by the appropriate inclusion of vertex corrections.\cite{VanishingSHE} However such cancelation  does not hold in the presence of magnetic fields and/or magnetic impurities.\cite{Inoe2,Pei-Wang} Furthermore, the vertex corrections are in fact identically zero in  2D-hole systems with {\it k}-cubic Rashba,\cite{Khaetskii}  2D $k$-cubic Dresselhaus electron systems, \cite{Malshukov} and Luttinger  Hamiltonians,\cite{Bernevig1,Schliemann1} for which the spin-Hall conductivity is shown to be robust against disorder. This holds however as long the standard definition of spin current is used.\cite{Comment1}  

A basic issue in spin transport theory is the use of an appropriate definition of spin current for systems in which the spin is not a conserved quantity. The standard definition of the spin current operator for spin-1/2 particles is the expectation value of the anticommutator of the velocity and spin operators, {\bf J}$_{s}=\langle \frac{\hbar}{4} \{ \sigma_z,   \hat{\bn v} \}\rangle$, where $\sigma_z$ is the Pauli spin $z-$component and  $\hat{\bn v}$   the electron velocity operator, respectively. This definition has the appealing form that resembles the usual charge-current operator, and in addition, for spin-polarized systems, it yields to the expected difference between the spin-up and spin-down charge currents. However, the conventional definition of spin current operator has a caveat, it is not conserved in systems with SOI, rendering it incomplete in describing a true ``spin-current".\cite{Murakami2,Culcer} 

Recently Shi {\it et al.} \cite{Shi,PZhang} introduced an alternative definition of spin current which circumvents the latter issue. The proposed definition for the effective spin current operator is described by the time derivative of the spin displacement operator, which for spin-1/2 carriers is simply given by ${\cal \hat J}^{sz}=\frac{\hbar}{2}\frac{d (\hat r \sigma_z)}{d t}$.  The alternative definition adds to the conventional part {\bf J}$_s$, a spin source term (torque dipole density {\bf P}$_{\tau}$) associated to the electron spin precessional motion. In this way the total effective spin current density can be defined as ${\cal J}_s=$ {\bf J}$_s+${\bf P}$_{\tau}$, with ${\cal J}_s=$Re$\Psi^\dagger {\cal   {\hat J}}^{sz}\Psi $.   Most important, such effective spin current definition straightforwardly satisfies the continuity equation $\frac{\partial S_z}{\partial t}+\nabla\cdot{\cal J}_s=0 $, with  $S_z=\frac{\hbar}{2}\Psi^\dagger\sigma_z\Psi$ describing the spin density. It is also possible to establish an Onsager relation between the spin transport coefficients and the mechanical or thermodynamical force-driven transport coefficients. In addition, it vanishes for localized orbitals, predicting the expected zero spin Hall conductivity for insulators.\cite{PZhang}

Using the conserved spin-current operator of Ref.[\onlinecite{Shi}], T.-W. Chen {\it et al.}\cite{TWChen1}  predicted that the dc value of the spin Hall conductivity for linear Rashba and Dresselhaus SOI models has an opposite sign to that obtained using the conventional definition. N. Sugimoto {\it et al.}\cite{Sugimoto} have also studied the SHE and the conditions for nonzero spin Hall currents using the conserved current definition for both, linear and cubic Rashba models. Subsequently, we explored (Ref.[\onlinecite{Wong}]) the spin Hall conductivity for 2D electron gases with competing {\it k}-linear Rashba and Dresselhaus SOI in the ac frequency regime. There we have highlighted the contrasting results when the conserved spin current operator is applied to 2DEGs. Most recently, T.-W. Chen {\it et al.}\cite{TWChen} have investigated the spin torque and spin Hall currents in generic 2D spin-orbit Hamiltonian $H_{so}=$A({\bf k})$\sigma_x -$B({\bf k})$\sigma_y$ also using the conserved spin-current operator and found that regardless of the detailed form of the energy dispersion [{\it i.e.} A({\bf k}) and B({\bf k}) coefficients], the conserved static ($\omega=0$) spin Hall conductivity changes its sign with respect to the conventional spin Hall conductivity.

In this paper, using the definition for the spin-current operator reported by Shi {\it et al.} \cite{Shi} we explore the behavior of the frequency dependent spin (Hall) $\sigma_{\mu\nu}^{sz}(\omega)$, and of the charge conductivity tensor $\sigma_{\mu\nu}^{ch}(\omega)$ for a 2DHG with {\it k}-cubic Rashba and Dresselhaus SOI. We show that the optical spectrum of the spin conductivity exhibit remarkable changes when this new definition of spin current is applied. A rather large response of the spin Hall conductivity is predicted to arise when using the conserved definition of spin current operator owing to a dominant contribution of the spin-torque term. In particular, we predict that the magnitude and sign of the dynamic spin current strongly depends on the electric field frequency as well as with the interplay of the Rashba and Dresselhaus spin-orbit coupling strengths. Such behavior is similar to that reported in 2DEGs,\cite{Wong} however, unlike the later case,  which gives vanishing SHE ($\sigma_{SH}(\omega)=0$) when the Rashba and Dresselhaus parameters have the same strength, a rather finite spin conductivity ($\sigma_{SH}(\omega)\neq0$) is obtained in general for the case of heavy-holes. Useful analytical expressions for the charge and spin (Hall) conductivities as a function of the frequency can be derived in the common limit $\varepsilon_{so}/\varepsilon_F \ll 1$,  being $\varepsilon_{so}$ the characteristic spin-orbit energy and $\varepsilon_F$ the Fermi energy of the heavy-holes. Interestingly, a straightforward connection between spin and charge conductivities can be established in such limit. 

The remaining of the present work is organized as follows. In Sec. II the Hamiltonian model for a 2DHG in the presence of both, $k$-cubic Rashba and Dresselhaus SOI is described. The main features of the frequency-dependent Kubo formula in linear response and its application to the spin-Hall conductivity are presented in Sec. III.  We offer a discussion on the connection between the spin and charge conductivities in Sec. IV. Section V is devoted for the discussion of the numerical results. We conclude in Sec. VI, and finally, in Appendixes A and B, we outline the derivation of the spin-current-charge-current correlation function and the longitudinal spin conductivity, respectively.  

\section{HAMILTONIAN MODEL}

We are interested to model spin (Hall) transport in 2DHGs formed in III-V semiconductor quantum wells with a relatively strong confinement potential ($\langle k^2_z\rangle >k^2$). In such systems, the resulting large splitting of heavy hole (HH) and light (LH) hole bands  are expected to yield fully filled LH states, as reported in recent experiments.\cite{Wunderlich} Hence the relevant contribution to the charge and spin conductivity at the Fermi energy arises from the HH bands alone. In such case we can describe the two-dimensional single particle (spin-3/2) hole system  by the effective Hamiltonian 
\begin{equation}\label{fullH}
H=\frac{\ p^2}{\ 2m^*}+H_{R}+H_D ,
\end{equation}
\noindent where $m^*$ is the effective mass for the HHs, and we have included the two most likely dominant SOI terms; namely, the $k$-cubic Rashba and Dresselhaus spin-orbit coupling for HHs.\cite{RWinkler,PrivateC-Winkler} In \eqref{fullH} the term $H_R$ denotes the Rashba spin orbit coupling  which arises from the structural inversion asymmetry (SIA) of the hole confining potential. It is given by\cite{Winkler}
\begin{equation}\label{R}
H_R=\frac{\ i\alpha}{\ \hbar^3}(\sigma_+p^3_--\sigma_-p^3_+),
\end{equation}

\noindent with, $p_{\pm}=p_x \pm i p_y$, being $p_{x,y}$ the components of the 2D momentum operator,   $\sigma_\pm = (\sigma_x \pm i \sigma_y)/2$, where $\sigma_{x,y}$ are the usual spin Pauli matrices, and $\alpha$ specifies the electrically tunable Rashba SOI parameter for HHs. Values  of the order of $10^{-22}$ eVcm$^3$ for $\alpha$ have been found in GaAs-AlGaAs quantum well samples with heavy-hole densities between $1.8\times 10^{10}$ cm$^{-2}$ to $4.2\times 10^{10}$ cm$^{-2}$.\cite{Winkler3}

The third term in \eqref{fullH}, $H_D$, is the Dresselhaus spin-orbit interaction for HHs which can be described by\cite{Bulaev}
\begin{equation}
\label{D}
H_D=-\frac{\ \beta}{\ \hbar^3}(\sigma_+p_-p_+p_-+\sigma_-p_+p_-p_+).
\end{equation}
\noindent  This term originates from the bulk-induced inversion asymmetry (BIA) of zincblende semiconductor structures. The spin-orbit parameter $\beta$ which is fix for a given system, can be estimated through the expression,  $\beta=3\gamma_0 \gamma \langle p_z^2 \rangle (E_g+E_{so})/E_{so}E_{hl}$, where $\gamma$ is the BIA (Dresselhaus) parameter, $\gamma_0$ is the Luttinger constant (within the spherical approximation, $\gamma_2\simeq \gamma_3=\gamma_0$). The energy parameters are as follows, $E_g$ is the fundamental gap, $E_{hl}$ and $E_{so}$  are the HH-LH energy gap and the  split-off--light-hole energy gap, respectively, all at the quantum well region. For a GaAs-based quantum well of 100 nm of width, $\beta$ is numerically estimated to be of the order of $10^{-22}$ eVcm$^3$.  The Dresselhaus term  \eqref{D} was earlier introduced by Bulaev and Loss\cite{Bulaev} in a study of spin relaxation and decoherence in quantum dots in perpendicular magnetic fields.  The importance of the $k$-cubic SOI related terms in the spin-splitting of the subband spectrum of HHs in quantum wells was first noticed by Rashba and Sherman. \cite{Sherman}

The eigenstates for the full Hamiltonian Eq.\eqref{fullH}  are  
\begin{equation}\label{eigenvectors}
\vert \psi_{k,\mu}(\textbf{r})\rangle =\frac{\ e^{i\textbf{k} \cdot \textbf{r}}}{\ \sqrt{2 A}} 
\begin{bmatrix} 1 \\ 
\mu e^{i(2\theta-\phi)} \\ \end{bmatrix}, 
\end{equation} 

\noindent with $\textbf{k}=(k_x,k_y)=(k\cos \theta, k \sin \theta)$ in polar coordinates, $\phi=\text {tan}^{-1}\frac{\ \alpha k_x-\beta k_y}{\ \alpha k_y-\beta k_x}$,  $A$ is the area of the system, and $\mu \in\{+1,-1\}$ denotes the pseudo-spin $\pm$3/2 HH branch. The energy spectrum is cubic in $k$ and it is given by  
\begin{align}\label{spectrum}
\varepsilon_\mu&=\frac{\ \hbar^2 k^2}{\ 2m^*}+\mu \Delta (\theta) k^3, \\
\Delta (\theta)&= \sqrt{\alpha^2+\beta^2-2\alpha \beta \sin(2\theta)}. 
\end{align}
\noindent Notice that, in spite of the obvious difference in the spin-orbit Hamiltonian between electrons (linear in $k$) and heavy-holes (cubic in $k$) in 2D systems, the angular anisotropy in the energy spectrum introduced by the expression $\Delta(\theta)$ acquires formally an identical algebraic form to that for electrons.\cite{Maytorena,Wong} Nevertheless, the $k$-cubic spin-splitting of the HH spin-branches, $\varepsilon_+(${\bf k}$)-\varepsilon_-(${\bf k}$)=2\Delta(\theta)k^3$, reveals already rather distinct features than the case for electrons as discussed below in more detail.

\begin{figure}  
\centerline{\includegraphics[width=2 in]{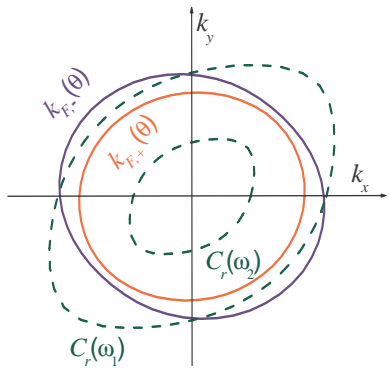}}
\caption{(Color online) Fermi contour $k_{F,\mu}(\theta)$ showing the angular anisotropy of the spin-split heavy-hole bands of a 2DHG with $k$-cubic Rashba and $k$-cubic Dresselhaus spin-orbit interaction. The dotted $C_r(\omega)$ curve results from the condition $\varepsilon_+(${\bf k}$)-\varepsilon_-(${\bf k}$)=\hbar\omega$, see text for details.}
\label{fig1}
\end{figure}

Following Schliemann and Loss,\cite{Schliemann} for a given Fermi energy $E_F$ and vanishing temperature, the two dispersion branches $\varepsilon_\mu(${\bf k}$)$ produces two (spin-dependent) Fermi wave numbers $k_{F,\mu}(\theta)$ which can be written compactly as
\begin{equation}
k_{F,\mu}(\theta)=\frac{\ k_F}{\ \sqrt{2}}\Bigr[\frac{\ \sqrt{4u-3u^2}-\mu u}{\ \sqrt{u(2-u)}}\Bigr],
\end{equation}

\noindent being $u(\theta)=1-\sqrt{1-2 \epsilon^2 \lambda(\theta)}$, with $\lambda(\theta)=\Delta^2(\theta)/(\alpha^2+\beta^2)$, $\epsilon=\varepsilon_{so}/\varepsilon_F$ and $\varepsilon_{so}=\sqrt{\alpha^2+\beta^2} k_F^3$ is the characteristic spin-orbit energy. \cite{Schliemann}The Fermi energy for vanishing spin-orbit coupling is estimated as $\varepsilon_F=\frac{\ \hbar^2k_F^2}{\ 2m^*}$, with Fermi wave number $k_F=\sqrt{2\pi n_h}$ and $n_h$ is the heavy-hole density concentration.

Now, in the presence of an oscillating electric field of frequency $\omega$ the allowed direct transitions between the HH spin-branches $\varepsilon_{\pm}(k,\theta)$ should satisfy the equation $\varepsilon_+(k,\theta)-\varepsilon_-(k,\theta)=\hbar\omega$, which for a fixed photon energy, defines a elliptical curve $C_r(\omega)$ in the rotated $\tilde k-$space described by the equation 
\begin{equation}
\frac{\tilde{k}^2_x (\tilde{k}^2_x+\tilde{k}^2_y)^2}{k^6_a}+\frac{\tilde{k}^2_y(\tilde{k}^2_x+\tilde{k}^2_y)^2}{ k^6_b}=1,
\end{equation} 

\noindent with semi-axis of lengths $k^3_a(\omega)=\hbar\omega/2|\alpha-\beta|$ and $k^3_b(\omega)=\hbar\omega/2|\alpha+\beta|$ oriented along the principal axes (1,1) and (-1,1) of the reference $k$-space (Fig. \ref{fig1}).  The symmetry points $k_{a,b}(\omega)$ are given by $k_a(\omega)=k_{F,-}(\pi/4)$ and $k_b(\omega)=k_{F,+}(3\pi/4)$. From this equations two characteristic energies are determined,
\begin{align} \label{funo}
\hbar\omega_a &=2|\alpha-\beta|k_{F}^3\Bigr[1+\frac{\ m^* k_F}{\ \hbar^2}|\alpha-\beta|\Bigr]^3,\\
\hbar\omega_b &=2|\alpha+\beta|k_{F}^3\Bigr[1-\frac{\ m^* k_F}{\ \hbar^2}|\alpha+\beta|\Bigr]^3.
\end{align}

\noindent additionally, another two absorption frequencies can be identified, namely,   
\begin{equation}\label{ftres}
\hbar\omega_{\pm}=2|\alpha\mp\beta|k_{F}^3\Bigr[1\mp\frac{\ m^* k_F}{\ \hbar^2}|\alpha\mp\beta|\Bigr]^3,
\end{equation}

\noindent which corresponds to transitions between states at the points $k_a(\omega)=k_{F,+}(\pi/4)$ and $k_b(\omega)=k_{F,-}(3\pi/4)$, respectively. These four characteristic frequencies above will play an important role in the overall response of the charge and spin (Hall) conductivities as we shall discuss later. 

\section{spin conductivity in the frequency domain}

Consider a weak and spatially homogeneous electric field $E(\omega)$\^{y} oscillating with frequency $\omega$  in the plane of the 2DHG. Within the linear response theory, the frequency-dependent spin conductivity  describing a $z$-polarized-spin current flowing in the $\nu=x,y$ direction can be characterized by the Kubo formula\cite{Schliemann}

\begin{multline}\label{kubo}
\sigma_{\nu y}^{sz}(\omega)=\frac{\ e}{\ \hbar A \tilde{\omega}} \int^\infty_{0} e^{i\tilde{\omega}t} \sum_{{\bf k},\mu}f(\varepsilon_\mu)_{T=0}  \langle \psi_{{\bf k},\mu}(\textbf{r}) \vert \\
\times [{\cal \hat J}_\nu^{sz}(t), \hat v_y (0)] \vert \psi_{{\bf k},\mu}(\textbf{r}) \rangle dt,
\end{multline}

\noindent where $e=-|e|$ is the hole electric charge. Here we have assumed noninteracting carriers with $f(\varepsilon_\mu)$  the Fermi-Dirac distribution function (in the limit of zero temperature) and $\tilde{\omega}=\omega+i\eta$. The vanishing parameter $\eta>0$   is just an artifact to regularize the integral and guarantee causality properties of the Kubo formula.\cite{Schliemann} However, phenomenologically it can be understood  as a measure of the hole momentum dissipation effects due to impurity scattering events, phonons or any other many-body interactions effects. In this way, $\eta^{-1}=\tau$ will represent here a finite life-time of the (spin-orbit coupled) quasi-particles due to scattering with holes.\cite{vertex} 

We remark here that instead of employing the conventional spin current operator in the Kubo formula, we have employed the alternative (conserved) spin current operator introduced by Shi {\it et al.}\cite{Shi} written in the interaction picture, ${\cal \hat J}_{\nu}^{sz}(t)$. The conserved spin current operator for HHs in the Schr\"odinger picture is defined by ${\cal \hat J}^{sz}_{\nu}=\frac{\ 3\hbar}{2}\frac{d (\hat r_{\nu} \sigma_z)}{d t}$, where the prefactor 3/2 comes from the projection of the total angular momentum of the HH states along the growth direction.\cite{PZhang} To obtain explicitly ${\cal \hat J}_{\nu}^{sz}(t)$ we  first take the $\nu$-component ($\nu=x,y$) of the current operator with a spin moment polarized along the $z$-axis at time $t=0$ (via Heisenberg equation of motion), yielding the operator ${\cal \hat J}_{\nu}^{sz}(0) ={\hat J}_{\nu}^{c}(0)+{\cal \hat J}_{\nu}^{\tau}(0)$, where 
\begin{equation}\label{Jc}
{ \hat J}_{\nu}^{c}(0)=\frac{\ 3\hbar}{\ 4}\left\{ \sigma_z,\frac{\ p_{\nu}}{\ m^*}\right\},
\end{equation}

\noindent is just the conventional spin-current operator definition, and
\begin{equation}\label{Jt}
{\cal \hat J}_{\nu}^{\tau}(0)=\frac{\ 3 \hbar}{\ 4}\left\{\sigma_y,\frac{{\cal P}_{\nu x}}{m^*}\right\}-\frac{\ 3 \hbar}{\ 4}\left\{\sigma_x,\frac{{\cal P}_{\nu y}}{m^*}\right\},
\end{equation}

\noindent represents the spin-torque contribution to the spin-current operator, being ${\cal P}_{\nu \nu^{\prime}}=\{\nu,\Gamma_{\nu^{\prime}}\}$ with  $\Gamma_{x}={m^*}[\alpha \,p_{y}(3p^2_{x}-p^2_{y})-\beta \,p_{x}p^2 ]/ \hbar^4$ and $\{,\}$ denoting the anticommutator. To get  $\Gamma_{y}$ simply replace $p_x\rightarrow p_y$ and $p_y\rightarrow p_x$ (see Appendix A). Note that in the absence of SOI, the operator ${\cal P}_{\nu,\nu^{\prime}}=0$, and the total spin-current operator reduces trivially  to the conventional form \eqref{Jc}. The effective spin current operator is then expressed in the interaction picture, ${\cal \hat J}_{\nu}^{sz}(t)=e^{iHt/\hbar}{\cal \hat J}_{\nu}^{sz}(0)e^{-iHt/\hbar}$ which allows us to calculate the spin-current--charge-current correlation function of the Kubo formula once the single-particle operator of the carrier velocity  $\hat v_y (0)$ is obtained (see Appendix A). After some straightforward algebraic manipulations, it is then possible to express the frequency dependent spin conductivity as the sum of two terms, namely
\begin{equation} \label{numint}
\sigma_{\nu y}^{sz}(\omega)=\sigma^c_{\nu y}(\omega)+\sigma^\tau_{\nu y}(\omega),
\end{equation}

\noindent where the first term to the right, $\sigma^c_{\nu y}(\omega)$,  comes from the conventional part of spin-current definition and it is given by
\begin{equation}\label{conv}
\sigma^c_{\nu y}(\omega)=-\frac{\ 3e}{\ 4\pi^2m^*}{\hspace{-0.1cm}}\int_0^{2\pi} {\hspace{-0.3cm}}d\theta g_{\nu}(\theta)  \frac{\ \cos \theta}{\ \Delta}\int_{k_{F,+} }^{k_{F,-} } {\hspace{-0.3cm}} dk\frac{\ k^4 }{\omega_k^2-\tilde{\omega}^2}
\end{equation}

\noindent with $g_{\nu}(\theta)=(\beta^2-\alpha^2-2\Delta(\theta)^2)(\delta_{\nu x} \cos\theta+\delta_{\nu y} \sin\theta)$ where $\delta_{\nu\nu^{\prime}}$ is the usual Kronecker delta and $\omega_k = 2\Delta k^3/\hbar$. The second term in \eqref{numint} arises from the spin-torque contribution to the net spin current and reads 
\begin{equation}\label{torq}
\sigma^\tau_{\nu y}(\omega)=\frac{\ 6e}{\ \hbar^2 \pi^2 m^*}{\hspace{-0.1cm}}\int_{0}^{2\pi}{\hspace{-0.3cm}} d\theta g_{\nu}(\theta)  \cos \theta \Delta {\hspace{-0.1cm}} \int_{k_{F,+} }^{k_{F,-} }{\hspace{-0.3cm}} dk\frac{\ k^{10}}{\ \Bigr(\omega_k^2-\tilde{\omega}^2\Bigr)^2}.
\end{equation}

The $k$-integrals in expressions \eqref{conv} and \eqref{torq} can be calculated exactly, however they lead to rather cumbersome expressions and shall not be given here. In general, for the case $\alpha,\beta\not=0$, the $\theta$ integrals can not be performed straightforwardly and a numerical integration has to be implemented. For a pure Rashba ($\beta=0$) or Dresselhaus ($\alpha=0$) system,  the angular dependence of the integrand above reduces significantly, leading {\it e.g.} to a vanishing longitudinal spin conductivity. Although the transverse spin (Hall) conductivity yields close analytic form for $\sigma^{c}_{xy}(\omega)$ and $\sigma^{\tau}_{xy}(\omega)$, the resulting expressions still somewhat complicated making almost impossible to retrieve any valuable physical insight from them. Thus, before going into the numerics, and in order to make progress understanding qualitatively the physics here, it is useful to consider first the behavior of the spin conductivity in the limit $\varepsilon_{so}/\varepsilon_F\ll 1$.    This is a reasonable limit which holds typically in 2DHG's in III-V based semiconductor heterostructures and for which analytical formulas of the frequency-dependent spin conductivity can be derived. From Eqs. \eqref{conv} and \eqref{torq}, we obtain to leading order in $\varepsilon_{so}/\varepsilon_F\equiv \epsilon$,
\begin{equation}\label{analytichh}
\sigma_{\nu y}^{sz,0}(\omega)\simeq\sigma^{c,0}_{\nu y}(\omega)+\sigma^{\tau,0}_{\nu y}(\omega) +{\cal O}(\epsilon^3),
\end{equation}

\noindent where the dominant contribution from the conventional part of the spin Hall ($\nu=x$) reads (see Appendix B for $\nu=y$),
\begin{equation}\label{analitica-convhh}
\frac{\ \sigma^{c,0}_{xy}(\omega)}{\ 9e/8\pi}   =    \frac{\ 1}{\ 3}\Bigr[2+ \frac{2\hbar^2\tilde{\omega}^2+\prod_{\mu}\xi_{\mu}}{\prod_{\mu}(\xi_{\mu}^2-\hbar^2\tilde{\omega}^2)^{1/2}}\Bigr ],
\end{equation} 
  
\noindent being  $\xi_{\mu}=2(\alpha+\mu \beta)k_F^3$, while leading contribution of the spin-torque can be written in the form
\begin{equation}\label{analitica-tauhh}
\frac{\ \sigma^{\tau,0}_{x y}(\omega)}{\ 9e/8\pi}  =  -2\frac{\ \sigma^{c,0}_{xy}(\omega)}{\ 9e/8\pi}-\frac{2}{3}{\cal G}(\omega) ,
\end{equation}
 \noindent where we have defined the auxiliary function,
\begin{widetext}
\begin{equation}
\label{auxiliar}
{\cal G}(\omega)   = \frac{\hbar^2\tilde\omega^2}{\prod_{\mu}(\xi_{\mu}^2-\hbar^2\tilde{\omega}^2)^{1/2}}\Bigr[2-\frac{(2\hbar^2\tilde{\omega}^2+\prod_{\mu}\xi_{\mu})  [\hbar^2\tilde{\omega}^2-4\varepsilon^2_{so}]}{ \prod_{\mu}(\xi_{\mu}^2-\hbar^2\tilde{\omega}^2)}\Bigr].
\end{equation}
\end{widetext}

It is illustrative to study the behavior of the spin-Hall conductivity in the static limit ($\omega=0$). In the presence of weak disorder, it must be analyzed comparing the characteristic spin-orbit energy $\varepsilon_{so}$ with the energy scale of the impurity scattering $\hbar\eta$ (related to the time-relaxation rate $\tau$ as $\eta=\tau^{-1}$) by taking into account both limiting cases, $\varepsilon_{so}\ll\hbar\eta$ and $\varepsilon_{so}\gg\hbar\eta$.  Assuming first that the impurity scattering dominates over the spin-orbit coupling, we expand eqs. \eqref{analitica-convhh} and \eqref{analitica-tauhh} in powers of $\varepsilon_{so}/\hbar\eta\ll1$. To lowest order we get

\begin{equation}
\frac{\ \sigma_{xy}^{c,0}(0)}{\ 9e/8\pi}\simeq \frac{\ 4}{\ 3}\Bigr[\frac{\ 2\varepsilon_{so}^2+\varepsilon_R^2-\varepsilon_D^2}{\ \hbar^2\eta^2}\Bigr],
\end{equation}

\noindent while $\sigma_{xy}^{\tau,0}(0)=0$, being $\varepsilon_R=\alpha k_F^3$ and $\varepsilon_D=\beta k_F^3$. 

On the other hand, if the impurity scattering is weak compared to the spin-orbit coupling ($\varepsilon_{so}/\hbar\eta\gg1$) we obtain to second order
\begin{equation}\label{esoetaconv}
\frac{\ \sigma_{xy}^{c,0}(0)}{\ 9e/8\pi}\simeq \frac{\ 2}{\ 3}+\frac{\ 1}{\ 3}\textrm{sgn}(\alpha^2-\beta^2)\Bigr[1-\frac{\ \hbar^2\eta^2}{\ \varepsilon_{so}^2}\frac{\ {\cal R}}{\ 4}\Bigr],
\end{equation}

\noindent with a nonzero torque part,

\begin{equation}\label{esoetatau}
\frac{\ \sigma_{xy}^{\tau,0}(0)}{\ 9e/8\pi}\simeq -\frac{\ 4}{\ 3}-\frac{\ 2}{\ 3}\textrm{sgn}(\alpha^2-\beta^2)\Bigr[1-\frac{\ \hbar^2\eta^2}{\ \varepsilon_{so}^2}\frac{\ {\cal R}}{\ 2}\Bigr],
\end{equation}

\noindent where we have defined the dimensionless parameter ${\cal R}=(\alpha^2+\beta^2)(3\alpha^2-\beta^2)/(\alpha^2-\beta^2)^2$.  The above expressions for $\sigma_{xy}^{c,0}(0)$ reduces, in each limit case, to the known formulas for $\beta=0$ (only cubic-Rashba) reported in Ref. [\onlinecite{Schliemann}].

Moreover, note that, at zero frequency and for ultra-clean  ($\eta\rightarrow 0^+$) samples, {\it i.e.} for ${\tilde\omega}\rightarrow 0+i0^+$ the function ${\cal G}(\omega)\rightarrow 0$, which in turn simplify  \eqref{analitica-tauhh} to $\sigma^{\tau,0}_{x y}(0)= -2\sigma^{c,0}_{x y}(0)$. Thus, the torque-dipole contribution  makes the total spin-conductivity, $\sigma^{sz,0}_{x y}(0)$ to change its sign with respect to the conventional result. It can be shown that this relationship is valid as well for the longitudinal spin conductivity, $\sigma^{\tau,0}_{y y}(\omega)$ in the same limit. From Eqns. \eqref{analytichh}, \eqref{esoetaconv} and \eqref{esoetatau} the static value of the spin Hall conductivity for the 2DHG (free of disorder) takes the universal form
\begin{equation}\label{dclimit}
\sigma_{xy}^{sz}(0)=\begin{cases}
-9e/8\pi  & \text{for $\alpha^2>\beta^2$},\\
-6e/8\pi  & \text{for $\alpha=\beta$},\\
-3e/8\pi  & \text{for $\alpha^2<\beta^2$}.
\end{cases}
\end{equation}

In contrast with the 2DEG case,\cite{Sinitsyn,Wong} Eq. \eqref{dclimit} predicts a non-vanishing dc spin-Hall conductivity ($-6e/8\pi$) when the spin-orbit parameters, $\alpha$ and $\beta$ have the same strength.  There is also an asymmetry of the dc spin Hall conductivity depending whether $\alpha^2>\beta^2$ or $\alpha^2<\beta^2$, unlike its counterpart for electrons in the clean limit. We obtain $\sigma_{xy}^{sz}(0)=-9e/8\pi$ for $\beta = 0$ whiles $\sigma_{xy}^{sz}(0)=-3e/8\pi$ for $\alpha = 0$, in striking difference to the analogous for electrons where $\pm e/8\pi$ is obtained, respectively.

It is known that for a 2DEG in presence of Rashba and (linear) Dresselhaus SOI, with spin-orbit parameters $\alpha_e$ and $\beta_e$, respectively, the operator ${\cal S}=(\sigma_x\pm\sigma_y)/\sqrt{2}$ provides an additional conserved quantity when $\alpha_e=\pm\beta_e$ since the  Rashba and Dresselhaus SOI Hamiltonians for electrons commutes with $\cal S$ in such case. \cite{symmetry} This symmetry leads to $k$-independent spin-states which in turn yields a suppression of the spin Hall effect at $\alpha_e=\pm\beta_e$. For a 2DHG with cubic Rashba and Dresselhaus SOI the scenario is completely different since such symmetry  is forbidden for $\alpha=\pm\beta$. Indeed, the operator ${\cal S}$ does not commute with the 2DHG-Hamiltonian with cubic Rashba and Dresselhaus SOI (incidentally this only occurs for $p_y=0$ which yields no spin Hall effect) leading in general to $k$-dependent spin states (see Eq. \eqref{eigenvectors}). Therefore a non-zero response is expected for the heavy-hole spin Hall conductivity even for $\alpha=\pm \beta$. Alternatively, for electrons the unitary transformation $\sigma_x \rightarrow -\sigma_y$, $\sigma_y \rightarrow -\sigma_x$ and $\sigma_z \rightarrow -\sigma_z$ makes the Rashba and Dresselhaus couplings to be interchanged, along with a  change of sign of the spin current, $J_s^z\rightarrow -J_s^z$.\cite{Shen04} This symmetries  explains why the spin Hall conductivity only differ by a sign depending upon the competitions of both strengths, as well as its suppression at the symmetry point $\alpha_e = \pm\beta_e $. However, for cubic-Rashba \eqref{R} and cubic-Dresselhaus \eqref{D} SOI there is no a mapping of one Hamiltonian to the other (the symmetry above for electrons is broken here), and thus, the dc spin Hall conductivity is asymmetric depending upon the relative ratio between $\alpha$ and $\beta$.  As we will see below, the ac spin Hall conductivity for heavy holes is also finite for $\alpha=\pm \beta$.
  
Furthermore, in the ac domain ($\omega\neq 0$) it can be demonstrated that the torque spin Hall conductivity \eqref{analitica-tauhh} is related to the conventional spin Hall conductivity \eqref{analitica-convhh} through the relation $\sigma^{\tau}_{x y}(\tilde{\omega})=-(2+\tilde{\omega}\partial/\partial\tilde{\omega})\sigma^{c}_{xy}(\tilde{\omega})$, in agreement with Ref.[\onlinecite{TWChen}]. This relation does not hold however for the longitudinal (along the driving field) spin conductivity. Moreover  a spin current can be generated along the driving field direction as long the system is submitted to a non-zero Dresselhaus SOI, as has been previously reported for electrons.\cite{Sinitsyn} 

\section{connection between charge and spin conductivities}

	In a 2DHG subject to an oscillating electric field, the driven electric current is described by the charge current conductivity tensor $\Sigma_{\nu y}(\omega)=\delta_{\nu y}\sigma_D(\omega)+\sigma^{ch}_{\nu y}(\omega)$, in which $\sigma_D(\omega)=\frac{ine^2}{\ m^*\tilde{\omega}}=\frac{\sigma_o}{1-i\omega\tau}$, is the dynamic Drude conductivity, being $\sigma_o=\frac{ne^2\tau}{\ m^*}$ the static Drude conductivity and $\sigma^{ch}_{\nu y}(\omega)$ is the contribution due to inter spin-split induced transitions. Within the linear-response Kubo formalism, following a similar procedure to previous section, the spin-orbit induced contribution to the charge current conductivity takes the form
\begin{equation}\label{chintnum}
\frac{\ \sigma^{ch}_{\nu y}(\omega)}{\ e^2/2\pi \hbar}=\frac{\ 2i}{\pi^2   \tilde{\omega}\hbar^3}\int_0^{2\pi}d\theta h_\nu(\theta) \frac{\cos\theta  }{\ \Delta}
\int_{k_{F,+} }^{k_{F,-} }dk \frac{\ k^8}{ \omega_k^2-\tilde{\omega}^2}
\end{equation}

\noindent where $h_\nu(\theta)=(\delta_{\nu x}\sin\theta-\delta_{\nu y}\cos\theta)(2\Delta^2+\alpha^2-\beta^2)^2$. Note that, as it occurs with the dynamic Drude conductivity, the model leads to a charge current conductivity which is singular for the simultaneously clean ($\eta\rightarrow 0$) and zero-frequency limit. 
Although for $\alpha,\beta\not=0$ the $k$-integrals in the above expression are elementary, the subsequent integrals in $\theta$ do not have an analytical solution. It is then illustrative to examine  the limit $\epsilon\ll 1$ to obtain analytical expressions of  $\sigma^{ch}_{\nu y}(\omega)$. It can be shown that for $\alpha$ and $\beta$ different from zero, the charge conductivity can be related to eqn. \eqref{analitica-convhh} for $\sigma^{c,0}_{xy}(\omega)$ and $\sigma^{c,0}_{yy}(\omega)$ (see Appendix B) as

\begin{eqnarray}\label{carga-spin}
\frac{\ \sigma^{c,0}_{xy}(\omega)}{\ 9e/8\pi}& =&   {\cal A}(\omega) \Bigr[\frac{\ i\sigma^{ch,0}_{yy}(\omega)}{\ e^2/2\pi\hbar}-\frac{\ (3\alpha^2+\beta^2)k_F^6}{\ \hbar\tilde{\omega}\varepsilon_F}\Bigr],\\
\frac{\ \sigma^{c,0}_{yy}(\omega)}{\ 9e/8\pi}& = &  {\cal A}(\omega)\Bigr[\frac{\ -i\sigma^{ch,0}_{xy}(\omega)}{\ e^2/2\pi\hbar}+\frac{\ 2\alpha\beta k_F^6}{\ \hbar\tilde{\omega}\varepsilon_F}\Bigr],
\end{eqnarray}

\noindent being ${\cal A}(\omega)=(4/3)\hbar\tilde{\omega}\varepsilon_F (2\hbar^2\tilde{\omega}^2+\prod_{\mu}\xi_{\mu})^{-1}$. From the equations above it follows that the longitudinal spin conductivity can be determined through the (transverse) charge Hall conductivity induced by the SOI. A similar situation occurs for the spin Hall conductivity, which in turn can be written in terms of the longitudinal charge conductivity. Such connection can be simply interpreted as a manifestation of the inverse SHE. It can also be viewed as a consequence of the existence of a charge imbalance in the sample due to a mean SOI induced transverse force in the (anisotropic) $k$-cubic Rashba-Dresselhaus hole system, as recently predicted to occur in such systems by T.-W. Chen {\it et al}.\cite{T.-W.Chen09} The latter implies the presence of a finite Hall voltage (transverse charge conductivity) in the absence of magnetic field and as long an electric field is present in the SOI system.  As the electric field $E(\omega)$ is assumed here along the $y$ direction, the total spin current is given by ${\cal J}^{sz}(\omega)=[\sigma^{sz}_{xy}(\omega)$\^{x}$+\sigma^{sz}_{yy}(\omega)$\^{y}$]E(\omega)$.  Thus from Eqs. \eqref{analitica-convhh}, \eqref{analitica-tauhh} and \eqref{carga-spin}, the spin conductivity tensor and the spin current can be obtained, in principle, through the measure of the charge conductivity  within the frequency domain alone. We believe that this may provide an electrical method to detect heavy holes spin currents in presence of $k$-cubic Rashba and Dresselhaus SOI. 

\section{NUMERICAL RESULTS AND DISCUSSION}

We  start our discussion of the numerical results by considering the isotropic case first, {\it i.e.} when only one type of SOI, Rashba ($\beta=0$) or Dresselhaus ($\alpha=0$), is present. We considered a 2DHG formed in a GaAs-AlGaAs quantum well with a heavy hole effective mass of $m^*=0.51\,m_o$ and a moderated sheet hole density of $n_h=3\times 10^{11}$ cm$^{-2}$; here $m_o$ is the free electron mass.  For this system, the Rashba parameter has been calculated\cite{Winkler2} to be $\alpha=7.48\times 10^{-23}$ eVcm$^3$, which gives rise to a HH spin-splitting at the Fermi energy of $\Delta_{R}=2\alpha k^3_F \simeq 0.38$ meV. The Fermi wave number for vanishing SOI is estimated from $k_F=\sqrt{2\pi n_h}$ . The parameter describing the momentum relaxation rate has been chosen such that $\hbar \eta =0.035$ meV, value that would corresponds to samples with mobilities $\mu\simeq e \tau/m^*\simeq 20$ m$^2$/Vs and relaxation times of $\tau\simeq 118$ ps. 

\begin{figure}
\hspace{-2 cm} 
\centerline{\includegraphics[width=2.5 in]{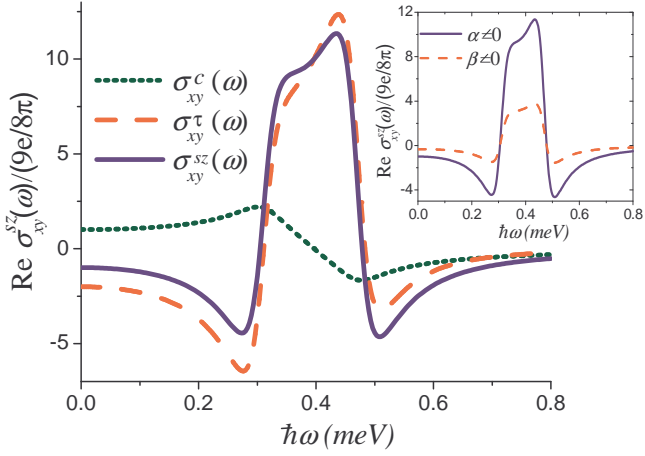}}
\caption{(Color online) Real part of the spin Hall conductivity as a function of the photon energy $\hbar\omega$ for a GaAs-AlGAs 2DHG system with Rashba SOI only ($\beta=0$). The dotted (green) curve is obtained using the conventional definition of spin-current, whereas the dashed (red) and solid (blue) curves are the torque contribution and total spin conductivity, respectively, employing the conserved spin-current operator. In the inset, the case with $\beta=0$ ($\alpha=7.48\times 10^{-23}$ eVcm$^3$) is contrasted with the $\alpha=0$ ($\beta=7.48\times 10^{-23}$ eVcm$^3$) case. See text for values of remaining parameter used. }
\label{fig2}
\end{figure}

In Fig. ~\ref{fig2} we plot the numerical integration of the real part of the spin-Hall conductivity against the frequency of the applied electric field, namely, expressions \eqref{numint}, \eqref{conv} and \eqref{torq} for the case with only Rashba SOI ($\beta=0$). The short-dashed (green) line depicts the conventional term of spin Hall conductivity, $\sigma^{c}_{xy}(\omega)$.  It shows a resonance-like behavior at the energy  $\hbar\omega_+\simeq 2\alpha k^3_{F,+}$ and a Fano-like behavior at $\hbar\omega_-\simeq 2\alpha k^3_{F,-}$, which from Eq.\,\eqref{ftres} they would correspond to the minimum and maximum photon energy required to induce optical transitions between the initial $\mu=+1$ and final $\mu=-1$ pseudo-spin-3/2 split branches (see Fig. ~\ref{fig1}). At low frequencies it approaches the universal value $9e/8\pi$, while it vanishes for high frequencies ($\omega >> \Delta_R/\hbar$). Because of the finite value of the damping parameter $\eta$, the spectrum has a somewhat smoothed shape.

The torque dipole contribution $\sigma^{\tau}_{xy}(\omega)$, large-dashed (red) lines, shows a rather different behavior. It develops a strong resonance at $\hbar \omega \simeq \Delta_R$ as well as near the frequencies $\omega_{\pm}$.  At low energies, the torque contribution approaches to the value $-2(9e/8\pi)$, and as a result, the total spin Hall conductivity changes its sign relative to the conventional result  reaching the value $-9e/8\pi$ in the static limit. In general, we notice that the frequency-dependent torque-dipole response dominates the shape of the spectrum of the total spin Hall conductivity and consequently, a dramatic change of the overall shape of the spectrum $\sigma_{xy}^{sz}(\omega)$ is observed. Such large effect of the inclusion of the torque term in the new definition of the spin current has been also reported to occur in 2DEG with Rashba and Dresselhaus SOI.\cite{Wong}
The inset of Fig. 2 shows a comparison of the spin conductivities between the case with $\beta=0$ ( $\alpha=7.48\times 10^{-22}$  eVcm$^3$), and the opposite case, {\it i.e.} $\alpha=0$ (with $\beta=7.48\times 10^{-22}$ eVcm$^3$) case. Qualitatively, they share essentially the same spectral features, but with a clear difference in the intensity for the central peak. In the dc limit, the effective spin Hall conductivity for the system with $\alpha\neq 0$ and $\beta=0$  converges to $-9e/8\pi$, whereas for the case with $\alpha= 0$ and $\beta\neq 0$ it reaches the constant value $-3e/8\pi$. Such asymmetric response is expected due to the formally non-equivalent $k$-cubic (Rashba and Dresselhaus) Hamiltonians, in contrast with the case of electrons, where the Rashba and Dresselhaus Hamiltonians can formally be mapped into each other, as discussed in section III.

\begin{figure}  
\centerline{\includegraphics[width=2.5 in]{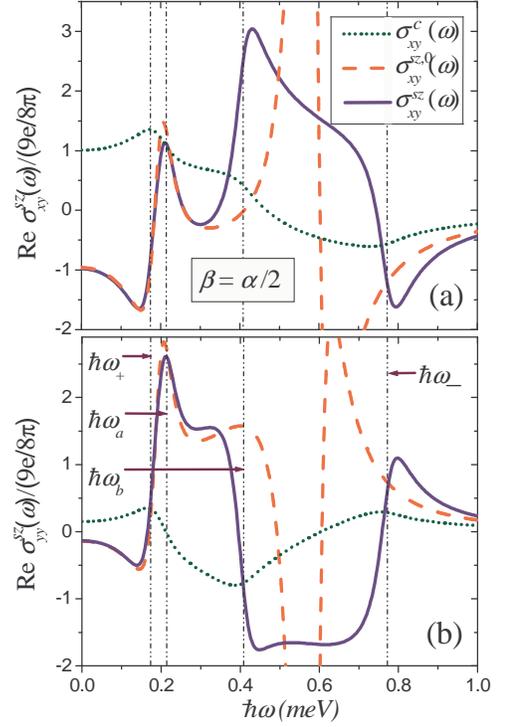}}
\caption{(Color online)   Frequency dependent spin (a) Hall and (b) longitudinal conductivity for a 2DHG system with finite Rashba and Dresselhaus SOI ($\beta = 0.5\alpha$). The dotted (green) curve shows the result using the standard definition of spin current, the case for the conserved spin-current operator ${\cal J}_{\nu y}^{sz}$ in the limit $\epsilon\ll1$ is shown in dashed (red) line, and the outcome of the exact numerical integration of Eq. \eqref{numint} is shown with solid (blue) lines. Here $\hbar\eta=0.035$ $meV$ and other parameters are as in Fig. \ref{fig2}. We identify four characteristic frequencies, two defining the optical absorption edges, $\omega_{-}$ and $\omega_{+}$, while the other two correspond to the peaks of the spin conductivity occuring at $\omega_{a}$ and $\omega_{b}$. The latter two arises due to the symmetry of the spin-split conduction bands in $k-$space at the Fermi level. }
\label{fig3}
\end{figure}

New interesting features appear when the interplay of the Rashba and Dresselhaus SOI is considered. In Fig.\,\ref{fig3} (a) and (b) we plot the frequency dependent spin Hall and longitudinal conductivity for  $\beta=0.5\alpha$  (the remaining parameters are as in Fig.\ref{fig2}). The results obtained through the approximated formula \eqref{analytichh} and the exact numerical integration of \eqref{numint} are presented for comparison, showing a good agreement, mostly at low frequencies. Here it is also evident the remarkable difference between the optical spectrum resulting from the use of the standard and the conserved spin-current operator definition, ${\cal J}^{s,z}_{\nu}$.  The physical origin of the main spectral features can be understood in terms of the anisotropic spin splitting caused by the simultaneous presence of the Rashba and Dresselhaus couplings, in  an analogous way to that of the 2DEG case. \cite{Maytorena,Wong}

In the limit of vanishing temperature, the sum over states in eq. \eqref{kubo} is restricted to the region between the Fermi contours $k_{F,+}(\theta)\leq k\leq k_{F,-}(\theta)$, for which $\varepsilon_+(k,\theta)\leq\varepsilon_F\leq\varepsilon_-(k,\theta)$.   The number of direct transitions that can take place at the energy $\hbar\omega$ only involve states with wave vectors that satisfy the equation $\varepsilon_+(k,\theta)-\varepsilon_-(k,\theta)=\hbar\omega$. As discussed in the introduction, four distinctive frequencies can be identified given the anisotropic $k-$space available for the optical response; namely, $\omega_a$, $\omega_b$ and $\omega_{\pm}$. 

The peaks observed in the optical conductivity $\sigma_{xy}^{sz}(\omega)$ of Fig.\,3 correspond to energy transitions $\hbar\omega_{a,b}$ involving states in the vicinity of the symmetry points $k_a(\omega)=k_{F,-}(\pi/4)$ and $k_b(\omega)=k_{F,+}(3\pi/4)$, respectively (see Eq. \eqref{funo}). Similarly, we can see that there are absorption edges at energies corresponding to transitions between states at the points $k_a(\omega)=k_{F,+}(\pi/4)$ and $k_b(\omega)=k_{F,-}(3\pi/4)$, which would correspond to the transition energies $\hbar\omega_{\pm}$,  respectively.  For clarity and to guide the eye, such characteristic energies are indicated in Fig.\,3 with dashed vertical lines. It can also be observed a large separation between resonances $\hbar \omega_b$ and $\hbar \omega_-$ unlike the 2DEG case.\cite{Wong} Surprisingly, such observation indicates  that there is a larger splitting between the Fermi contours $k_{F,+}$ and $k_{F,-}$ along the (-1,1) direction for 2DHGs than for 2DEGs.   The finite value of $\hbar \eta$ chosen here, besides of introducing  an overall smoothing of the spectrum, it also yields to a slight shifting of the appearance of the peaks in relation to these (four) characteristic energies.  It is clear that the definition of spin-current operator by Shi {\it et al.} \cite{Shi} yields a drastic different frequency response from that predicted by the conventional definition. In addition,  as it occurs with the pure Rashba (or Dresselhaus) SOI case, the torque dipole contribution turns out to be the dominant term in the spin-Hall conductivity.
 
We have also explored the effect induced of varying the ratio $\alpha/\beta$ on the conserved spin conductivity as a function of the exciting frequency. In particular, in Fig.\,4 (a) the spin Hall response is shown for the specific values of $\alpha/\beta = 0,0.5,1$ and $1.5$ while fixing the rest of the sample parameters as in Fig.\,3. Notice that the energy separation of the resonance peaks becomes larger as the aspect ratio $\alpha/\beta$ is increased.  A non-zero ac spin Hall conductivity is observed in general for the $\alpha=\beta$ case, in contrast to the 2DEG which is zero at all frequencies. It is also appreciated that the resonance frequencies $\omega_+$ and $\omega_b$ disappear for this case. This can be explained in terms of the collapse of the equations \eqref{funo} and \eqref{ftres} at $\alpha=\beta$.  The latter is due the overlapping of the spin $\pm3/2$ dispersion laws, $\varepsilon_{+}(${\bf k}$)=\varepsilon_{-}(${\bf k}$)$, at $k_{F\pm}=\pi/4$ and at $k_{F\pm}=3\pi/4$, which occurs precisely at the symmetry point $\alpha=\beta$. Such effect is emphasized in Fig. 4(b) where a color map of the spin Hall conductivity is plotted as a function of a continuous variation of the ratio $\alpha/\beta$ and the exciting frequency.

\begin{figure} 
\centerline{\includegraphics[width=2.5 in]{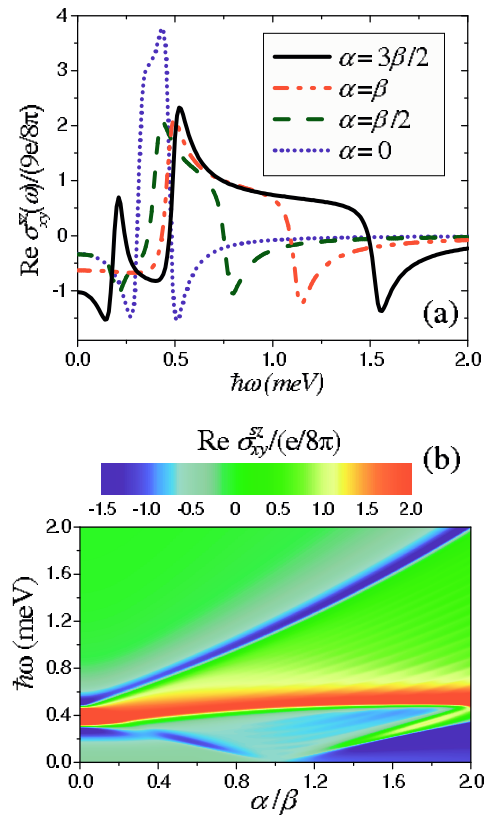}}
\caption{(Color online) (a) Total spin Hall conductivity for different aspect ratios of $\alpha/\beta$ calculated employing the  conserved spin-current operator ${\cal J}^{sz}$.   The parameters are as in Fig. 2. Notice that the resonance peaks tend to separate in energy as the ratio $\alpha/\beta$ is increased. (b) Color contour map of the ac spin Hall conductivity showing its behavior with a continuous variation of the frequency $\omega$ and to the relative ratio $\alpha/\beta$. }
\label{fig4}
\end{figure}

\begin{figure}  
\centerline{\includegraphics[width=2.5 in]{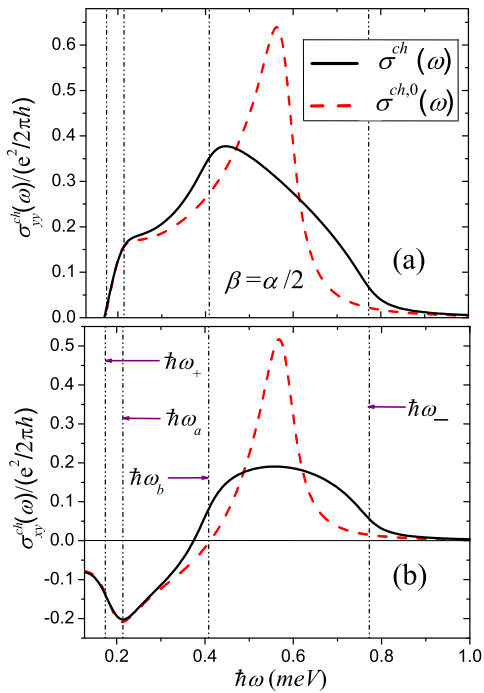}}
\caption{(Color online) Frequency dependent (a) longitudinal and (b) transverse spin orbit contribution to the charge conductivity for a 2DHG system with $k$-cubic Rashba and Dresselhaus SOI ($\beta = 0.5\alpha$). The dashed curves shows the result in the limit $\epsilon\ll1$ whiles the solid curves corresponds to the exact numerical integration of Eq.\,\eqref{chintnum}.  There is a good agreement between both cases for frequencies lower than $\omega_b$.  The sample parameters are the same as in Fig \ref{fig2}. The four main frequencies $\omega_{+}$, $\omega_{a}$, $\omega_{b}$ and $\omega_{-}$ are shown as vertical lines to guide eye.}
\label{fig5}
\end{figure}

In figure 5 we have plotted the (a) longitudinal ($\nu=y$) and (b) transversal ($\nu=x$) spin-orbit contribution to the charge conductivity for a 2DHG system with $\beta = 0.5\alpha$. The dashed curve corresponds to the plot with expressions in the limit $\varepsilon_{so}/\varepsilon_F\ll1$, whiles the solid line represent those obtained by exact numerical integration of Eqn. \eqref{chintnum}. The results shows a quite good agreement between both curves for frequencies smaller than $\omega_b$. The four main frequencies $\omega_{+}$, $\omega_{a}$, $\omega_{b}$ and $\omega_{-}$ are shown here as vertical lines. Interestingly, we have noticed that these distinctive resonance frequencies, can in principle, be used to estimate the strengths of the spin-orbit parameters $\alpha$ and $\beta$. This is done as follows. From Eqs.\,\eqref{funo} and \eqref{ftres} assuming $\varepsilon_{so}/\varepsilon_F\ll1$ and $\alpha,\beta>0$ we arrive to the useful expressions 
\begin{align}\label{almbe}
\alpha\Theta(\alpha-\beta)+\beta\Theta(\beta-\alpha)&=\frac{\ \omega_++\omega_a+\omega_b+\omega_-}{\ 8\hbar^{-1}k_F^3},\\
\label{almbe2}
\beta\Theta(\alpha-\beta)+\alpha\Theta(\beta-\alpha)&=\frac{\ \hbar k_F^{-1}}{\ 6m^* }  \frac{\ \omega_++\omega_a-\omega_b-\omega_-}{\ \omega_++\omega_a+\omega_b+\omega_-},
\end{align}
\noindent where $\Theta(x)$ is the usual Heaviside step function. Assuming that such characteristic frequencies can be experimentally identified, either via the spin-Hall conductivity or from the charge conductivity tensor measurements, thus the values for both, the Rashba and Dresselhaus coupling parameters can simultaneously be estimated from the formulas above. First it is needed to retrieve if the system has $\alpha<\beta$ or $\alpha>\beta$. This can be done by varying a gate voltage to change the heavy-hole concentration $n_h$, which consequently changes continously the tunable $\alpha$ parameter.\cite{Winkler2} Since $\beta$ does not depend on the carrier density, then the plot of the right hand side of Eq.\,\eqref{almbe} should give basically a constant output, which incidentally will coincide with value of $\beta$ if the 2DHG has $\alpha<\beta$, see Fig.\,6. Then the value of $\alpha$ is determined through the expression  in Eq.\,\eqref{almbe2}. On the other hand, if a linear behavior of \eqref{almbe} with $n_h$ is observed,   this is an indicative that $\alpha>\beta$ and its value gives an estimate for $\alpha$ for a given $n_h$ concentration. Once $\alpha$ is determined in this regime, the value of the constant parameter $\beta$ can be readily obtained from Eq.\,\eqref{almbe2}. 

\begin{figure}  
\centerline{\includegraphics[width=2.4 in]{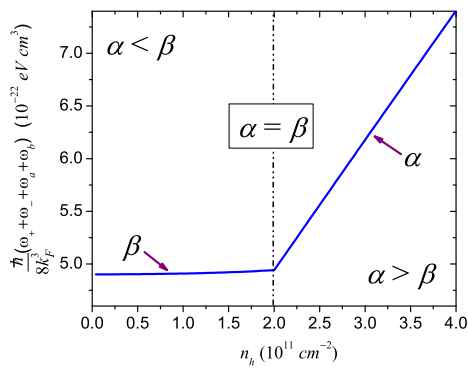}}
\caption{(Color online) Plot of the sum of the frequencies $\omega_+$, $\omega_a$, $\omega_b$, and $\omega_-$ in units of $8\hbar^{-1}k_F^3$  as a function of the heavy hole  concentration, $n_h$. For $\alpha>\beta$ a linear variation with $n_h$ is observed, while for the opposite case ($\beta>\alpha$) an almost uniform behavior is seen in a wide range of hole concentrations.  The latter responds to the fact that typically $\alpha$ is dependent on the heavy holes density, while $\beta$ is not. The dramatic difference of the behavior between the cases $\alpha>\beta$ and $\beta>\alpha$ as the density is varied can in principle be used to experimentally estimate the value of the coupling parameters. See text for details.}
\label{fig6}
\end{figure}

\section{Conclusions}

In this paper we have examined the spin and charge conductivity in the frequency regime for a 2D heavy hole gas with $k$-cubic Rashba and Dresselhaus SOI employing a recently proposed spin current operator.  Our results shows that the optical spectrum of the spin (Hall) conductivity changes substantially when the conserved spin current operator is used due to a significant contribution of the spin torque term. In the dc limit we find that spin Hall conductivity for a pure $k$-cubic Rashba is 3 times the value obtained for the pure $k$-cubic Dresselhaus system, being $\sigma^{s,z}_{xy}(0)=-9e/8\pi$ and $\sigma^{s,z}_{xy}(0)=-3e/8\pi$, respectively. Such anisotropy (symmetry breaking) is understood in terms of the absence of mapping the Rashba $\leftrightarrow$ Dresselhaus Hamiltonians, in contrast with the case for electrons. It is argued that such asymmetry it is also responsible for the non-vanishing dc spin Hall conductivity ($\sigma^{s,z}_{xy}(0)=-6e/8\pi$) when the spin-orbit Hamiltonians, Rashba and Dresselhaus have the same coupling strength.  In the ac limit, it is shown that the magnitude and the sign of the dynamic spin current is rather sensitive to the frequency and to the spin-orbit ($\alpha$ and $\beta$) coupling strengths. The latter suggests the possibility of the optical manipulation of spin currents in addition to the control obtained through external bias. The dramatic differences of the spin Hall conductivity response predicted here with the use of the new conserved spin-current operator in relation to the calculated with its conventional counterpart, it is clearly a call for the experimentalist to perform measurements in the frequency domain and validate (or discard) its applicability in describing spin-transport. In addition, we have shown that the angular anisotropy of the spin-splitting energy induced by the interplay between the Rashba and Dresselhaus couplings gives rise to four characteristic resonance frequencies, which in principle, can be employed to estimate the strengths of the spin-orbit parameters through optical spectroscopy and/or transport measurements.

\section{ACKNOWLEDGMENTS} We are thankful to R. Winkler for useful comments. This work was supported in part by DGAPA-UNAM IN113-807-3 and by the Mexican Council for Science and Technology (CONACyT).

\appendix
\section{Spin-current-charge-current correlation function}

In this appendix we briefly outline the derivation of the expectation value for spin-current-charge-current correlation function  $\langle \psi_{{\bf k},\mu} (\textbf{r}) \vert [\hat{{\cal J}}_x^{S,z}(t), \hat {v}_y (0)] \vert \psi_{{\bf k},\mu}(\textbf{r}) \rangle\equiv{\cal F}_{\mu}({\bf k},t)$ appearing in the Kubo formula Eq.\,\eqref{kubo}. We begin by writing the carrier velocity operator via the Heisenberg equation of motion $  {\hat {\bn v}}(0)=\frac{i}{ \hbar}[ H, {\hat {\bn r}}]$. For the $\nu = y$ component, the velocity operator reads,
\begin{equation}
\hat v_y(0)=\frac{1}{\ m^*}\left(p_y+\sigma_x \frac{\partial \Gamma_x }{\partial p_y}+\sigma_y\frac{\partial \Gamma_y}{\partial p_y}\right),
\end{equation} 

\noindent where we have defined,
\begin{equation}
\label{Gamma}
\Gamma_{x}=\frac{m^*}{ \hbar^4}\left[\alpha \,p_{y}(3p^2_{x}-p^2_{y})-\beta \,p_{x}p^2 \right].
\end{equation}

To get  $\Gamma_{y}$ simply replace  $p_x\rightarrow p_y$ and $p_y\rightarrow p_x$ in \eqref{Gamma}. It follows trivially that in the absence of SOI, the operator $\Gamma_{x,y}=0$, and the total velocity operator reduces to the dispersionless form $p_y/m^*$. Next we proceed to calculate the operators ${x}(t)$, $\sigma_x (t)$ and $\sigma_y (t)$ in the interaction picture, ${\hat {\cal O}}(t)= e^{iHt/ \hbar}{\hat{\cal O}}(0)e^{-iHt/ \hbar}$. After some algebraic manipulations, the position operator reads,
\begin{equation}
\begin{split}
{x}(t)&={x}(0)+\frac{t}{m^*}\Bigr( p_x+\sigma_x\frac{\partial \Gamma_x}{\partial p_x}+\sigma_y\frac{\partial \Gamma_y}{\partial p_x}\Bigr)\\&\quad+\frac{\ \hbar p_y}{\ 2p^2}\frac{\ (\beta^2-\alpha^2-2\Delta^2)}{\ \Delta^2} {\cal K}(t), 
\end{split}
\end{equation}
\noindent whiles the spin Pauli operators in the interaction picture take the form,
\begin{align}
\sigma_x(t)&=\sigma_x(0)+ \Sigma_x(t),&
\sigma_y(t)&=\sigma_y(0)- \Sigma_y(t),&
\end{align}
\noindent here we have introduced the time-dependent operators 
\begin{align}
\Sigma_x(t) &= m^*\Gamma_y {\cal L}(t)\\
\Sigma_y(t) &= m^*\Gamma_x {\cal L}(t)
\end{align}

\noindent in which
\begin{equation}
\begin{split}
{\cal K}(t)&=\frac{\ 2}{\ \hbar\omega_p}[\omega_pt-\sin(\omega_pt)]\,{\cal R}_{xy}-2\sin^2(\omega_pt/2 )\,\sigma_z,\\
{\cal L}(t)&=\frac{\ 2}{\ \hbar\omega_p}\Bigr\{\sin(\omega_pt)\sigma_z-\frac{\ 1}{\ \hbar\omega_p} {\cal R}_{xy}\sin^2(\omega_pt/2 ) 
        \Bigr\},
\end{split}
\end{equation}

\noindent with ${\cal R}_{xy}=m^*(\sigma_x \Gamma_y-\sigma_y \Gamma_x)$ and $\omega_p=2\Delta p^3/\hbar^4$. Note that  $\sigma_x$, $\sigma_y$, $p$, $p_x$ and $p_y$ are  all given in the Schr\"odinger picture. The expressions above are needed in the calculation of the commutator $[\hat{\cal J}_x^{s,z}(t),\hat{v}_y(0)]$, in which the effective spin-current operator is written in the interaction picture, {\it i.e.} ${\hat {\cal J}}_x^{s,z}(t)= e^{iHt/ \hbar}{\hat{\cal J}}_x^{s,z}(0)e^{-iHt/ \hbar}$. At $t=0$ we have ${\hat {\cal J}}_x^{s,z}(0) = {\hat J}_x^{c}(0)+{\hat{\cal J}}_x^{\tau}(0)$, as given explicitly by Eqs.\, \eqref{Jc} and \eqref{Jt}. Then the commutator of the velocity operator with the conventional spin current reads,
\begin{equation}
\begin{split}
\label{cJc}
[\hat{J}_x^{c}(t),\hat{v}_y(0)]  & = \frac{\ 3\hbar}{\ 4}\left[e^{iHt/\hbar}\left\{ \sigma_z,\frac{\ p_{\nu}}{\ m^*}\right\}e^{-iHt/\hbar} ,\hat{v}_y(0)\right]\\
  & =  \frac{\ 3\hbar}{\ 2} \frac{\ p_x}{\ m^*}\left[  \sigma_{z}(t), \frac{1}{m^*}\left(\sigma_x\frac{\partial \Gamma_x}{\partial p_y}+\sigma_y \frac{ \partial \Gamma_y}{\partial p_y}\right)\right]\, , 
\end{split}
\end{equation}

\noindent whereas the torque spin-current operator yields
\begin{equation}\label{cJt}
\begin{split}
[\hat {\cal J}_x^{\tau}(t),\hat v_y(0)]& =\frac{\ 3 \hbar}{\ 4}\left[e^{iHt/\hbar}\left\{\sigma_y,\frac{{\cal P}_{xx}}{m^*}\right\}e^{-iHt/\hbar},\hat{v}_y(0) \right] \\ & - \frac{\ 3 \hbar}{\ 4}\left[e^{iHt/\hbar}\left\{\sigma_x,\frac{{\cal P}_{\nu y}}{m^*}\right\}e^{-iHt/\hbar},\hat v_y(0)\right],
\end{split}
\end{equation}

\noindent with ${\cal P}_{\nu \nu^{\prime}}=\{\nu,\Gamma_{\nu^{\prime}}\}$. After some algebraic manipulations we arrive to,
\begin{equation}
\begin{split}
\label{cJt}
[\hat{\cal J}_x^{\tau}(t),\hat v_y(0)]&= \frac{\ 3 \hbar}{\ 2}\sum_{i}^{3}\sum_j^2 \left( \left[\hat{x}{\cal M}_i,{\cal N}_j\right]+\left[{\cal M}_i\hat{x},{\cal N}_j\right]\right)\\&+\frac{\ 3\hbar}{\ 2}\frac{\ 2p_x}{\ m^*}t \sum_{i}^{3}\sum_j^2[{\cal M}_i,{\cal N}_j]
\end{split}
\end{equation}

\noindent with $\hat{x}=\hat{x}(0)$ and the remainder operators are

\begin{align}
{\cal N}_1&=\frac{1}{m^*}\frac{\partial \Gamma_x}{\partial p_y}\,\sigma_x\, ,& {\cal N}_2&=\frac{1}{m^*}\frac{\partial \Gamma_y}{\partial p_y}\,\sigma_y\, , \\
{\cal M}_1&=\frac{1}{m^*}\Gamma_y\cos(\omega_pt)\,\sigma_x\,, &  
{\cal M}_2&=-\frac{1}{m^*}\Gamma_x\cos(\omega_pt)\,\sigma_y\,, \\
{\cal M}_3&=-\frac{\ \hbar\omega_p}{\ 2}\sin(\omega_pt)\,\sigma_z\, .&  
\end{align}

Adding up  the expectation value of \eqref{cJc} and \eqref{cJt} with the aid of \eqref{eigenvectors}, we finally  arrive to a explicit formula for the spin-current-charge-current function ${\cal F}_{\mu}({\bf k},t)$, which is given by

\begin{multline}
\label{valoresperado}
{\cal F}_{\mu}({\bf k},t)= 6i\frac{\ (k_x^2-k_y^2)k^2(2\alpha^2+\Delta^2)}{\ \hbar}\cos(
\omega_k t)\\+3i\mu \frac{\hbar k_x^2 k 
\left (\beta^2-\alpha^2-2\Delta^2\right )}{m^*\Delta} \Bigr [ \cos(
\omega_k t)
- \omega_k\,t\sin(\omega_k t)\Bigr ],
\end{multline}

\noindent with $\omega_k = 2\Delta k^3/\hbar$ and $\mu=\pm$ the pseudo-spin $\pm 3/2$. The result of Eq.\,\eqref{valoresperado} together with the factor $e^{i\tilde{\omega}t}$ in  Eq.\,\eqref{kubo} leads expressions exactly integrable in time for the conventional \eqref{conv} and torque \eqref{torq} contribution to the effective spin Hall effect.

\section{Longitudinal spin current}

In this appendix we write down useful analytical expressions for the ac and dc longitudinal spin conductivity, $\sigma^{sz}_{yy}(\omega)$. Expanding  Eqs. \eqref{conv} and \eqref{torq} to leading order in $\varepsilon_{so}/\varepsilon_F$ by assuming $\varepsilon_{so}/\varepsilon_F<<1$ for $\alpha$ and $\beta$ different from zero, the conventional part of the longitudinal ($\nu=y$) spin conductivity takes the form

\begin{equation} \label{convyy}
\begin{split} 
\frac{\ \sigma^{c,0}_{yy}(\omega)}{\ -9e/8\pi}& =  \frac{\ 1}{\ 3}\left[1+\frac{\hbar^2\tilde{\omega}^2-4\varepsilon^2_{so}}{\prod_{\mu}(\xi_{\mu}^2-\hbar^2\tilde{\omega}^2)^{1/2}} \right] \\& \quad \times \left[\frac{2\hbar^2\tilde{\omega}^2+\prod_{\mu}\xi_{\mu}}{\ 8\alpha\beta k_F^6}\right ].
\end{split}
\end{equation} 
  
\noindent The spin-torque contribution reads
\begin{equation} \label{torqyy}
\frac{\ \sigma^{\tau,0}_{y y}(\omega)}{\ 9e/8\pi}  = -4\frac{\ \sigma^{c,0}_{yy}(\omega)}{\ 9e/8\pi}- \frac{4}{3}{\cal H}(\omega).
\end{equation}

\noindent with the function ${\cal H}(\omega)$ is defined as

\begin{equation}
\begin{split}
\label{auxiliardos}
{\cal H}(\omega)  & = \frac{\ (\alpha^2-\beta^2)}{\ 4 \alpha \beta}\left[1-\frac{\hbar^2\tilde{\omega}^2-4\varepsilon_{so}^2}{\prod_{\mu}(\xi_{\mu}^2-\hbar^2\tilde{\omega}^2)^{1/2}}\right]\\ & \quad + \frac{\ 4\alpha\beta \hbar^2\tilde{\omega}^2[\prod_{\mu}\xi_{\mu}+2\hbar^2\tilde{\omega}^2]k_F^6}{\prod_{\mu}(\xi_{\mu}^2-\hbar^2\tilde{\omega}^2)^{3/2}}.
\end{split}
\end{equation}

Clearly, for the case of identical SOI coupling strengths ($\alpha=\beta$) the function $\cal H (\omega)$ vanishes at all frequencies and $
 \sigma^{\tau,0}_{y y}(\omega) = -4 \sigma^{c,0}_{yy}(\omega)$. It can be shown that using Eqs.\,\eqref{convyy}, \eqref{torqyy} and \eqref{numint} within the dc limit and for ultra-clean ($\eta \rightarrow0^+$) samples, the static value of the longitudinal spin conductivity will reduce to
\begin{equation}
\sigma_{yy}^{sz}(0)=\begin{cases}
-\frac{\ 3e}{\ 8\pi}\frac{\ \beta}{\ \alpha} & \text{for $\alpha^2>\beta^2$},\\
0& \text{for $\alpha =\beta $},\\
\frac{\ 3e}{\ 8\pi}\frac{\ \alpha}{\ \beta}& \text{for $\alpha^2<\beta^2$}.
\end{cases}
\end{equation}

\noindent which depends on the ratio $\alpha/\beta$, as it occurs for a 2DEG.\cite{Sinitsyn}


\begin{thebibliography}{99}

\bibitem{Zutic}
I. Zutic, J. Fabian, and S. D. Sarma, Rev. Mod. Phys. {\bf 76}, 323 (2004).

\bibitem{Dyakonov}
M.I. Dyakonov, V.I. Perel, Sov. Phys. JETP (1971) 467.

\bibitem{Hirsch}
J. E. Hirsch, Phys. Rev. Lett. {\bf 83}, 1834 (1999).

\bibitem{Zhang}
S. Zhang, Phys. Rev. Lett. {\bf 85}, 393 (2000).

\bibitem{Hankiewicz-Vignale} E. M. Hankiewicz and G. Vignale, Phys. Rev. Lett. {\bf 100}, 026602 (2008).

\bibitem{Murakami} S. Murakami, N. Nagaosa, and S.-C. Zhang, Science {\bf 301}, 1348 (2003).

\bibitem{Sinova}
J. Sinova, D. Culcer, Q. Niu, N. A. Sinitsyn, T. Jungwirth, and A. H. MacDonald, Phys. Rev. Lett. {\bf 92}, 126603 (2004). 

\bibitem{Kato}
Y. K. Kato, R. C. Myers, A. C. Gossard, and D. D. Awschalom, Science {\bf 306}, 1910 (2004).

\bibitem{Stern}
N. P. Stern, S. Ghosh, G. Xiang, M. Zhu, N. Samarth, and D. D. Awschalom, Phys. Rev. Lett. {\bf 97}, 126603 (2006).

\bibitem{Wunderlich}
J. Wunderlich, B. Kaestner, J. Sinova, and T. Jungwirth, Phys. Rev. Lett. {\bf 94}, 047204 (2005).

\bibitem{Saitoh} E. Saitoh, M. Ueda, H. Miyajima, and G. Tatara, Appl. Phys. Lett. {\bf 88}, 182509 (2006).

\bibitem{Kimura} T. Kimura, Y. Otani, T. Sato, S. Takahashi, and S. Maekawa, Phys. Rev. Lett. {\bf 98} 156601 (2007).

\bibitem{Valenzuela}
S. O. Valenzuela and M. Tinkham, Nature (London) {\bf 442}, 176 (2006).

\bibitem{Ando} K. Ando, Y. Kajiwara, S. Takahashi, S. Meakawa, K. Takemoto, M. Takatsu and E. Saitoh, Phys. Rev. B {\bf 78}, 014413 (2008).

\bibitem{Seki}
T. Seki, Y. Hasegawa, S. Mitani, S. Takahashi., H. Imamura, S. Maekawa J. Nitta and K. Takanashi, Nature Mater. {\bf 7}, 125 (2008).

\bibitem{Bruene} C. Br\"une, A. Roth, E.G. Novik, M. K\"onig, H. Buhmann, E.M. Hankiewicz, W. Hanke, J. Sinova, and L. W. Molenkamp, arXiv:0812.3768, (2008). 

\bibitem{VanishingSHE}
J. Sinova, S. Murakami, S.-Q. Shen, and M.-S. Choi, Solid State Comm. {\bf 138}, 214 (2006); J.I. Inoue, G.E.W. Bauer, and L.W. Molenkamp, Phys. Rev. B {\bf 70}, 041303(R) (2004); E.G. Mishchenko, A.V. Shytov, and B.I. Halperin, Phys. Rev. Lett. {\bf 93}, 226602 (2004); O. Chalaev and D. Loss, Phys. Rev. B {\bf 71}, 245318 (2005); O.V. Dimitrova, Phys. Rev. B {\bf 71}, 245327 (2005).

\bibitem{Inoe2}
J.I. Inoue, T. Kato, Y. Ishikawa, H. Itoh, G.E.W. Bauer, and L. W. Molenkamp, Phys. Rev. Lett. {\bf 97}, 046604 (2006).

\bibitem{Pei-Wang} 
P. Wang, Y.-Q. Li, and X. Zhao, Phys. Rev. B {\bf 75}, 075326 (2007).

\bibitem{Khaetskii} 
A. Khaetskii, Phys. Rev. B {\bf 73}, 115323 (2006).

\bibitem{Malshukov}
A.G. Mal\'shukov and K. A. Chao, Phys. Rev. B {\bf 71}, 121308(R) (2005).

\bibitem{Bernevig1} 
B. A. Bernevig and S.-C. Zhang, Phys. Rev. Lett. {\bf 95}, 016801 (2005).

\bibitem{Schliemann1} 
J. Schliemann and D. Loss, Phys. Rev. B {\bf 69}, 165315 (2004).

\bibitem{Comment1} N. Sugimoto {\it et al.} (Ref.[\onlinecite{Sugimoto}]) have shown (within the Keldysh formalism) that the use of the conserved spin-current definition (Ref.[\onlinecite{Shi}]) and the proper account of the vertex corrections leads to a vanishing spin Hall conductivity for the cubic-Rashba model when considering $\delta$-function impurity potentials. This holds also for short range impurity scatters up to first order in the Born approximation.

\bibitem{Murakami2}
S. Murakami, N. Nagaosa, and S.C. Zhang, Phys. Rev. B {\bf 69} 235206(2004).

\bibitem{Culcer}
D. Culcer, J. Sinova, N.A. Sinitsyn, T. Jungwirth, A.H. MacDonald, and Q. Niu, Phys. Rev. Lett. {\bf 93}, 046602 (2004).

\bibitem{Shi}
J. Shi, P. Zhang, D. Xiao, and Q. Niu, Phys. Rev. Lett. {\bf 96}, 076604 (2006).

\bibitem{PZhang}
P. Zhang, Z. Wang, J. Shi, D. Xiao and Qian Niu, Phys. Rev. B {\bf 77}, 075304 (2008). 

\bibitem{TWChen1}
T.-W. Chen, C. M. Huang, and G. Y. Guo, Phys. Rev. B {\bf 73}, 235309 (2006).

\bibitem{Sugimoto}
N. Sugimoto, S. Onoda, S. Murakami and N. Nagaosa, Phys. Rev. B {\bf 73}, 113305 (2006).

\bibitem{Wong}
A. Wong, J. A. Maytorena, C. L\'opez-Bastidas, and F. Mireles, Phys. Rev. B {\bf 77}, 035304 (2008).

\bibitem{TWChen}
T.-W. Chen and G.-Y. Guo, Phys. Rev. B {\bf 79}, 125301 (2009).

\bibitem{RWinkler} R. Winkler, private communication.

\bibitem{PrivateC-Winkler} There is an additional linear in momentum SOI term (Ref.[\onlinecite{RWinkler}]) that contributes to the spin-splitting and arises as well due to BIA, however its contribution is typically much smaller than the cubic term for large hole densities ($n_s \gtrsim 1\times10^{11}$cm$^{-2}$) and will be neglected in this work.  

\bibitem{Winkler}
R. Winkler, Phys. Rev. B {\bf 62}, 4245 (2000)

\bibitem{Winkler3}
R. Winkler, H. Noh, E. Tutuc, and M. Shayegan, Phys. Rev. B {\bf 65}, 155303 (2002).

\bibitem{Bulaev}
D. V. Bulaev and D. Loss, Phys. Rev. Lett. {\bf 95}, 076805 (2005).

\bibitem{Sherman}
E. I. Rashba and E. Ya. Sherman, Phys. Lett. A {\bf 129}, 175 (1988).

\bibitem{Maytorena}
J. A. Maytorena, C. L\'opez-Bastidas, and F. Mireles, Phys. Rev. B {\bf 74}, 235313 (2006).

\bibitem{Schliemann}
J. Schliemann and D. Loss, Phys. Rev. B {\bf 71}, 085308 (2005).  

\bibitem{vertex} 
Vertex corrections are expected to contribute to the spin Hall current, mostly in the zero frequency  limit, however such effects are beyond the scope of this paper.

\bibitem{Sinitsyn}
N. A. Sinitsyn, E. M. Hankiewicz, W. Teizer, and J. Sinova, Phys. Rev. B {\bf 70}, 081312(R) (2004).

\bibitem{symmetry}
J. Schliemann, J. Carlos Egues, and D. Loss, Phys. Rev. Lett. {\bf 90}, 146801  (2003).

\bibitem{Shen04}
S.-Q. Shen, Phys. Rev. B {\bf 70}, 081311 (2004).

\bibitem{T.-W.Chen09} 
T.-W Chen, H.-C. Hsu, and G.-Y. Guo, Phys. Rev. B {\bf 80}, 165302 (2009).

\bibitem{Winkler2} 
{\it Spin-Orbit Coupling Effects in Two-Dimensional Electron and Hole Systems}, by R. Winkler (Springer-Verlag, Berlin, 2003).

\end{thebibliography}
\end{document}